\documentclass[12pt]{article}
\usepackage{authblk}
\usepackage{bm}
\begin {document}

\title {{Poincar{\'e} pressure and vorticity monopoles in the kinetic model of matter-extension}}

\author[a,b] {I. {\'E}. Bulyzhenkov}

\affil [a] {Moscow Institute of Physics and Technology,  
    } 

\affil[b] {Lebedev Physics Institute RAS, Moscow, 119991}


\date{}






\maketitle


{\it Local vorticity accompanies measurable distributions of Cartesian matter-extension in accordance with its geodesics for eddy current densities. Local 3-currents in the geodesic distribution of nonlocal mass-energy are responsible for the adaptive  Poincar{\'e} pressure to stabilize the radial cloud of the isolated continuous electron. The mechano-magnetic monopole of distributed vorticity is behind local displacements of material space and measurable densities of probing mechanic / electric charges. The non-local organization of mass-energy integral and local measurements of continuous inertial distributions correspond to different physical processes and equations. Eastern and Western dialectical teachings can be consistent in one monistic theory of kinetic space-matter for nonlocal mass-energy with local stresses.
}

\bigskip

{\bf Keywords}: geodesic distributions, mass-charge nonlocality, static vorticity, monopole, kinetic monism


\bigskip {\bf MSC2020 }: 83D05, 53C22, 81V10 

\section {Introduction}

In 1931, Dirac {\cite {Dir}} proposed the first magnetic monopole model to restore the expected symmetry of electric and magnetic charges in advanced physical theories. Later, t\rq{}Hooft \cite {Hoo}, Polyakov \cite {Pol} and other theorists predicted topological solutions for magnetic monopoles in promising ways to unify the known forces. Superconducting contours with SQUID detectors are main instruments in the experimental search for magnetic monopoles or its traces \cite {Dus}.
The MoEDAL experiment at the LHC was uniquely designed to find highly ionizing messengers of new physics such as magnetic monopoles \cite {Mod}. But how to explain that modern measurements are stubbornly silent, despite loud theoretical justification?

The textbook theory of relativistic fields traditionally  operates with point masses / charges in empty space of Newton. This classical concept of point matter cannot resolve the Newtonian / Coulomb energy divergence of radial fields. The alternative matter-extension \cite {Des, Mie, Ein} should avoid this divergence at the very center of continuous masses or charges. But here a new challenge arises. The extended electron must somehow stabilize the Coulomb repulsion of the distributed electric densities due to the negative Poincar{\'e} pressure \cite {Poi}. This inward pressure preventing an elementary charge cloud to disintegrate under its own Coulomb interaction. 
The physical origin of the complementary artifact of Poincar{\'e}\rq{s} pressure was not entirely clear even after its analytical calculations in the energy-momentum tensor of the distributed electric charge \cite {BulT}. The poorly discussed nature of Coulomb and Poincar{\'e} counterbalances in the extended electron stops its consideration in the university courses of physics. 
Below we look at these counterbalances in quantitative terms of Maxwell\rq{s} electrodynamics to understand the under-appreciated Cartesian physics with the partner organization of the continuous charge and its radial monopole of vortex densities.

Similarly to electrodynamics, the internal pressure must somehow prevent the gravitational collapse of the static mass density 
$ \mu^\prime = m r_o / 4 \pi r^2 (r + r_o)^2 $, analytically derived \cite {Bul} for the non-empty metric solution $ g_{oo} = r^2 / (r + r_o)^2, r_o = mc^2 / \varphi^2_{\!_G}$, $\varphi_{\!_G} \equiv c^2 / {\sqrt G} = 1.04 \times 10^{27} V $, in the co-moving system of references $r^\prime \equiv |{\bf x}^\prime| = |{\bf x}| \equiv r$. Below, our methodological approach to Cartesian matter-extension will clarify the vortex origin of the putative Poincar{\'e} pressure  in equilibrium organization of charge-energy distributions. 

The problem of undetectable magnetic monopoles should be better analyzed in the mechanical domain, where the geodesic law of General Relativity $ \mu^\prime c^2 u^\nu \nabla_\nu u_\mu \equiv \mu^\prime c^2 u^\nu (\nabla_\nu u_\mu - \nabla_\mu u_\nu) =
\mu^\prime c^2 u^\nu (\partial_\nu u_\mu - \partial_\mu u_\nu) = 0 $ well describes the measurable local motion of the probe body with the scalar mass density $ \mu^\prime \neq 0$ in an already established external field. But this 1913 law for local measurements and laboratory observations of probe bodies does not describe the prior distribution of the accelerating field through its nonlocal energy self-governance. The gravitating densities of the external mass-energy integral  $mc^2=const$ around its center of inertia can be nonlocally coordinated by the adaptive Poincar{\'e} pressure of the magneto-mechanic monopoles beyond the laws of local measurements.

For probe densities and measurements of their velocities, accelerating fields locally represent an external system of distributed mass-energy with the non-local charge integral $ {\sqrt G} m \equiv q_m = const $. Non-equilibrium material densities of non-empty space can move toward  their equilibrium shapes according to the Cartesian worldview of continuous matter-extension and the monistic statement of Descartes that invisible local motion is everywhere behind the observed mechanical events. Regardless of what and how one  measures locally, the scalar mass density $ \mu^\prime (x^\prime) $ in the co-moving frame $ \{ x^{\prime 0}/c; x^{\prime i} \}$ of nonlocally organized matter-extension 
can be monistically defined through the four-velocity $g^{\mu\nu}u_\mu = u^\nu \equiv dx^\nu / {\sqrt {g_{\rho\lambda} dx^\rho dx^\lambda }}$ 
or the Cartesian four-potential $U^\mu \equiv \varphi_{\!_G} u^\mu \equiv c^2u^\mu / {\sqrt G}  $ in the tensor mechanical fields $f^{\mu\nu} \equiv (\nabla^\mu U^\nu - \nabla^\nu U^\mu)$ and  $f_{\mu\nu} \equiv (\nabla_\mu U_\nu - \nabla_\nu U_\mu) = \varphi_{\!_G} (\partial_\mu u_\nu - \partial_\nu u_\mu)$.  Recall that the laboratory mass density $\mu(x) = \gamma(x) \mu^\prime (x^\prime) $ of the moving scalar density $\mu^\prime$ depends on the Lorentz factor $\gamma$ in the tensor formalism of relativistic physics.  Emphasizing the primacy of local observations or local measurements in the physical laboratory, we will relate below the immeasurable Poincare{\'e} pressure of local vorticity in the nonlocal organization of paired extended charges and continuous monopoles.

\section {New geodesics for inertial densities of nonlocally organized mass-energy} 

The density of the accelerating force for charged masses in electromagnetic fields, $ F_{\mu\nu}J^\nu/c = \mu^\prime c^2 du_\mu/dx^o$, is the well-known  contraction of the electric four-current $J^\mu \equiv 
c\nabla_\nu F^{\nu\mu}/4\pi$ with the electromagnetic field tensor $F_{\mu\nu} \equiv \nabla_\mu A_\nu - \nabla_\nu A_\mu$ \cite {Lan}. 
Thus, Maxwell electrodynamics suggests zero electric, $ F_{\mu\nu} J^\nu/c = 0$, and mechanical, $f_{\mu\nu}j^\nu/c = 0$, self-forces for  geodesic states in an isolated continuous electron. These densities of Lorentz 4-forces can vanish together in the monistic matter-extension of Descartes due to  the local relation $A^\mu(x) = const \ U^\mu (x) $ for elementary 4-potentials of electron\rq{s} densities in the absence of external influences. 

At first glance, the four mechanical equations $j^\nu f_{\mu\nu}/c = 0$ seem duplicate the 1913 geodesic equations $u^\nu \nabla_\nu u_\mu = 0$ under the equality $u^\nu \nabla_\mu u_\nu \equiv \nabla_\mu (u^\nu u_\nu /2) \equiv 0$ because  $j^\nu \propto u^\mu$ can be used for probe bodies. However, local measurements of material fields by probe bodies and nonlocal self-organizations of there fields before measurements are different physical processes. For the adaptive dynamics of nonlocal mass-energy distributions, one should not formally identify the divergence-free mechanical current $j^\nu (x)\equiv c \nabla_\lambda f^{\nu\lambda}(x)/4\pi $ with the translational four-current ${\sqrt G} \mu^\prime c u^\nu (x) $ of the inertial charge density  ${\sqrt G} \mu^\prime$. Cartesian vortex mechanics for the extended mass-energy $\varphi_{\!_G}\int {\sqrt {|g_{ij}|}} {\sqrt G} \mu^\prime (x^\prime)d^3 x^\prime =  mc^2 \equiv r_o \varphi^2_{\!_G}$ does not correspond to Newtonian translations of constant masses, and we adhere to the vortex structure of local currents in the spatial organization of nonlocal mass-energy:
\begin {equation}
   \frac {1}{c}{f_{\mu\nu}(x)  } { j^\nu (x)}  \equiv  \frac { \varphi^2_{\!_G}} {4\pi{\sqrt {-g} } }(\nabla_\mu u_\nu - \nabla_\nu u_\mu)
	\partial_\lambda [{\sqrt {-g} } g^{\alpha \nu} g^{\beta \lambda} (\nabla_\alpha u_\beta - \nabla_\beta u_\alpha)]   = 0. 
\end {equation}

Again, the GR acceleration $a_\mu \equiv c^2 D u_\mu/D s \equiv c^2 u^\nu \nabla_\nu  u_\mu = c^2 u^\nu (\partial_\nu u_\mu - \partial_\mu u_\nu)$ of observable masses in external metric fields can be derived from (1) only for the model  of collinear currents and velocities, $j^\mu \propto u^\mu$ of probe mass densities. The self-consistent metric dynamics of nonlocal energy distributions with divergence-free currents, $\nabla_\mu j^\mu \equiv 0$, in the Lorentz-type force density (1) does not follow Newton\rq{s} model of non-changing bodies in empty space gravitation. The geodesic law (1) for consistent densities of nonlocal mass-energy describes metric dynamics of continuous mechanical currents in a line of the Mie theory of electric matter \cite {Mie} and the Einstein-Infeld directive \cite {Ein} to a pure field theory of inertial spaces.

 It is quite difficult to find mathematical solutions of (1) with respect to $g_{\mu\nu}$ even for the symmetric affine connections in $\nabla_\mu u_\nu \equiv \partial_\mu u_\nu - {\Gamma_{\mu\nu}^\lambda u_\lambda } $ when $\nabla_\mu u_\nu - \nabla_\nu u_\mu  = \partial_\mu u_\nu - \partial_\nu u_\mu$. Fortunately for Cartesian physics,  high concentrations of geometricized matter-extension comply with the Euclidean 3-section in the pseudo-Riemann 4-interval, $ds^2 \equiv (c d\tau/ \gamma)^2  = [({g_{oo}} dx^o + g_{oi} dx^i )^2/g_{oo}]   - [(g_{oi} g_{oj}/g_{oo}) - g_{ij}]dx^idx^i $ $
= c^2 d^2\tau  (1 - \delta_{ij}dx^idx^j / c^2 d^2\tau )$, due to inherent metric symmetries \cite {Bul}. Ultimately, we tend to derive from the static equilibrium in (1) the already known \cite{Bul} field solutions for continuously distributed mass-energy 
$m \equiv r_o c^2/G $: 
\begin {eqnarray}
\cases {
0\leq g_{oo} = 1/g^{oo} = r^2/(r+r_o)^2 < 1, \cr
g_{oi} = g^{oi} = 0, \ g_{ij} = - \delta_{ij},  \ g^{ij} = - \delta^{ij}, \ {\sqrt {{-g}}} = {\sqrt {g_{oo}}}, \cr 
u^o \equiv dx^o/ds = \gamma / {\sqrt {g_{oo}}}, \ u_o \equiv g_{o\nu} dx^\nu/ds = \gamma {\sqrt {g_{oo}}}, \cr
u^i \equiv dx^i/ds =\gamma \beta^i, \ u_i \equiv g_{i\nu}dx^\nu/ds =  - \gamma \beta_i , 
\gamma = 1/{\sqrt {1 - \beta_i\beta^i}}, \cr  
\beta^i = \delta^{ij}\beta_j = dx^i/c d\tau = dx^i /c {\sqrt {g_{oo}}} dx^o , \ c d\tau \equiv {{g_{o\nu}dx^\nu } }/{\sqrt {g_{oo}}}   =  {\sqrt {g_{oo}}} dx^o.
}
\end {eqnarray}

A vorticity monopole in Euclidean material 3-space (due to the inherent metric symmetries $(g_{oi} g_{oj}/g_{oo}) - g_{ij} = \delta_{ij}$) will allow to associate the stabilizing Poincar{\'e} pressure with a non-zero 3-current in all spatial points. Let us first return to the local observation of probe bodies, the measurable replacement of which in metric fields can be described by the four-flow $i^\nu_{GR}(x) = \mu^\prime(x^\prime)  cu^\nu (x)$  of probe mass density ${\sqrt {g_{oo}}}i_{_{GR}}^o(x) = \mu^\prime(x^\prime) \gamma (x) =\mu (x)$ in the lab frame of references. The 1913 geodesic equation $ u^\nu (\partial_\mu u_\nu - \partial_\nu u_\mu) = 0$ reads for spatial translations of laboratory mass-energy densities $\mu^\prime \gamma c^2 = \mu c^2$ as 
\begin {eqnarray}
\cases {
 -\mu c^2  \beta^j(\partial_j \gamma {\sqrt {g_{oo}}} + \partial_o \gamma \beta_j\!) = 0 
\ $ for relativistic Bernoulli lines, $ \cr\cr
\mu c^2 \! {\sqrt {g^{oo}}}  (\!\partial_i \gamma \!{\sqrt {g_{oo}}}\! + \!\partial_o \gamma \beta_i)\!  = \!\mu c^2 \beta^j
(\partial_i \gamma \beta_j\! -\! \partial_j \gamma \beta_i)\! \equiv\! %
\mu c^2    [{\bm\beta} \! \times\! curl \gamma {\bm \beta} ]_i \cr\cr $ for ideal Euler flows without external forces, $ \cr\cr 
\partial_o  {\bm\Omega} = 
curl  [ 
{\bm \beta}\! \times \!{\bm \Omega}/{\sqrt {g^{oo}}}]  \
  $ for local vorticity $  
{\bm \Omega}\equiv  curl \gamma { \bm \beta}, \ div {\bm \Omega} \equiv 0.\cr
}
\end{eqnarray}
These relativistic generalization of the Euler - Bernoulli equations for ideal fluids describes the GR metric motion of non-metric  densities $\mu$ in the absence of radiation exchanges and frictions. But only dissipative processes can ultimately lead to the equilibrium distribution of nonlocal mass-energy.    
Below, a steady distribution with local vorticity ${\bm\Omega}\neq 0$ and $\partial_o  {\bm\Omega} \equiv \partial  {\bm\Omega}/ \partial x^o$ $ =0$ will be associated with a  vorticity-magnetic monopole that accompanies the creation of elementary masses and electric charges.

Now let us analyze the adaptive geodesic balance $j^\nu f_{\mu\nu}/c = 0$ of 4-force densities in (1) and all four components of the divergence-free mechanical current $j^{\mu}(x) \equiv c \nabla_\nu f^{\mu \nu}(x)/4\pi $ for the metric organization of nonlocal space-matter at 
the realistic 3-space geometry  with $ {\sqrt {-g}} = {\sqrt {g_{oo}}} =  1/{\sqrt {g^{oo}}}  $ and $g^{oi}  = 0$:     
\begin {eqnarray}
\cases {
j^\nu f_{o\nu}\! \equiv \! j^k f_{ok} \equiv {\bm j} {\bm E}_m \!= \!0, E^m_k\!  \equiv\! {\sqrt {g_{oo}}} D^m_k \equiv  f_{ok} \!\equiv\!
 - \varphi_{\!_G} (\partial_o \gamma \beta_k +  \partial_k u_o), \cr\cr
j^\nu f_{i\nu} \equiv j^of_{io} + j^k f_{ik} \equiv - j^oE^m_i   - [{\bm j}\times {{\bm B}_m}]_i = 0    , 
  \cr\cr
4\pi {\sqrt {g_{oo}}} j^o / c \equiv   \partial_j [{\sqrt {g_{oo}}} (g^{o\lambda}g^{j\nu} f_{\lambda\nu})] =  
- \delta^{jk}   \partial_j ( {f_{ok}}/{\sqrt {g_{oo}}} ) = - \partial_j D_m^j\cr
 \cr
4\pi {\sqrt {g_{oo}}} j^i / c \equiv    \partial_o [{\sqrt {g_{oo}}} (g^{i\lambda}g^{o\nu} f_{\lambda\nu})] + 
 \partial_k [{\sqrt {g_{oo}}} (g^{i\lambda}g^{k\nu} f_{\lambda\nu})] \cr\cr 
=   \partial_o D_m^i - 
\delta^{is} \varphi_{\!_G} \partial^p {\sqrt {g_{oo}}}  (\partial_s \gamma \beta_p - \partial_p \gamma \beta_s ) =
 \partial_o D_m^i  - [{\bm \partial} \times  {\bm H}_m ]^i\cr\cr
\partial_i f_{jo}\! +\! \partial_j f_{oi} \!+\! \partial_o f_{ij}\! \equiv 0 , \ \ curl {\bm E}_m \!\equiv\! - \partial_o {\bm B}_m,  
\cr\cr    
{{\bm B}_m} \equiv  \varphi_{\!_G}  curl \gamma { \bm \beta} \equiv \varphi_{\!_G} {\bm \Omega},  div {\bm B}_m \equiv 0,  {\bm H}_m\equiv {\sqrt {g_{oo}}} {\bm B}_m,
  {\bm D }_m  \equiv {\sqrt {g_{oo}}}{\bm E}_m . 
}
\end{eqnarray}
These mechanical equations resemble the Maxwell-Lorentz system with ${\bm B}_m$ $ \rightarrow $ ${\bm B}_e$ and $ {\bm H }_m  \rightarrow {\bm H}_e$. The opposite (mechanical) sign of the material densities $j^o$ and ${\bf j}$ is consistent  with different signs in the Poisson equation for Newton and Coulomb fields.  Now we look in (4) at the local geodesic balance 
$ {\bm E}_mj^o/c = - {\bm j} \times {\bm B}_m /c = {\bm B}_m  \times  {\bm j} /c$ of inertial material space under its nonlocal organization under the constant energy integral:
\begin {equation}
 - {   \varphi^2_{\!_G} (\partial_o \gamma {\bm \beta} +  {\bm \partial} u_o)  } 
   {\bm {\partial}} \left ( 
		\frac {  \partial_o \gamma {\bm \beta} +  {\bm \partial} u_o  }  { 4\pi\sqrt {g_{oo}}} \right)  =   
  {{\bm B}_m} \times \frac {	 \left [-curl {\bm H}_m +  {\partial_o {\bm D}_m}  \right]} { 4\pi {\sqrt {g_{oo}}}} .
			\end {equation}

Such local geodesics of variable material densities and their elastic tensions in non-equilibrium distributions of nonlocal mass-energy differs from the GR geodesic balance (3) for non-metric mass densities in external fields. 
The local force balance (5) with the vortex 3-currents of inertial space  can be applied to unperturbed self-governance of nonlocal mass-energy in the absence of wave exchanges.  The velocity-based distributions in the Maxwell-Einstein  system (1)-(5) can be equally applied to self-assembling of local mass flows and electric currents in chemical structures, as well as to the nonlocally correlated densities in an almost isolated galaxy. The right-hand side of the  geodesic balance  (5) represents the Poincar{\'e} pressure of immeasurable vortex currents, which compensate Newton/Coulomb force densities in stationary and static distributions of continuous matter.

\section {Space-matter vorticity for 3-displacements and mass/charge densities}

To apply the  geodesic balance (5) of Newton/Coulomb force densities and  counter-forces of the vorticity monopole (creating 
Poincar{\'e} pressure) to the simplest organization of nonlocal energy, we consider an isolated continuous electron. Its static  equilibrium should clarify the vorticity nature of the Poincar{\'e} pressure and the vortex origin of the measurable mass integral $m_o = 9,1 \times 10^{-31}{kg}$ (or the charge integral $|q_o| = 1,6\times 10^{-19} C $). We are interested in quantitative understanding  of how a continuous electron can maintain the stationary distribution of its continuous mass instead of the classical gravitational collapse (or Coulomb repulsive scattering).     


The higher-order derivatives in $<{\bm B}_m \times curl {\bm H}_m>_{x^o} \neq 0$ and the vortex self-governance of Cartesian matter-extension 
explains the existence of stationary and static strong metric fields by the non-zero 3-current densities at the right-hand side of (5).  
Does static mass-energy organization mean that the GR geodesic law $c^2u^\mu\nabla_\mu u_\nu = 0$ is incorrect for probe bodies? No, Einstein\rq{s} General Relativity self-consistently maintains the bouncing geodesic motion of probe bodies in strong static fields in the absence of inelastic energy exchanges \cite {BulA}.  And such cyclic motion of probe masses in static metric fields is consistent with their time-averaged equilibrium  $<g_{oo}({\bf x}, t)>_{x^o} = g_{oo}({\bf x})$ instead of an alleged collapse of material densities.

The stationary equilibrium of the time-averaged functions $F = \left< F\right>_{x^o}$, with $\partial_o F = 0$  in (5),   associates the  mechanical charge density  ${ \sqrt {g_{oo}}} j^o /c = -  div {\bm D}_m/4\pi > 0$  (where  $ {\bm D}_m  \equiv {\bm E}_m/{\sqrt {g_{oo}}}\equiv - \varphi_{\!_G} \gamma{\bm \partial} ln{\sqrt {g_{oo}}} 
 -   \varphi_{\!_G} {\bm \partial} \gamma \equiv \varphi_{\!_G}\gamma {\bm d}_m - \varphi_{\!_G} {\bm \partial} \gamma 
   $ is 3-vector displacement of material space) with  local vorticity $curl \gamma{\bm \beta} \equiv {\bm B}_m {\sqrt G}/ c^2 \neq 0$ in the Cartesian matter-extension:
\begin {eqnarray}
     \cases {
		{\bm E}_m (-div  {\bm D}_m  /4\pi {\sqrt {g_{oo}}}) = - {\bm B}_m  \times  curl {\bm H}_m/4\pi {\sqrt {g_{oo}}} \cr \equiv  
				({ B}_m^2 / 4 \pi ) {\bm d}_m  -({\bm B}_m \cdot {\bm d}_m / 4\pi){\bm B}_m - [{\bm B}_m  \times curl {\bm B}_m/4\pi] 
		\cr
		$ or Newton/Coulomb force density$ \ = \ $Poincar{\'e} counter-action$,   
		 \cr  \cr
		{\bm E}_m  \cdot{\bm B}_m = 0, \ {\bm E}_m \cdot  curl {\bm H}_m / {\sqrt {g_{oo}}}  = 0, \ 		       div ({\bm H}_m/{\sqrt {g_{oo}}}) = 0, \cr  \cr
				{\bm E}_m \times [{\bm B}_m  \times   curl {\bm H}_m] = {\bm E}_m \times \left [{\bm B}_m  \times   curl {\bm B}_m   
				+ {\bm B}_m  \times   [{\bm B}_m \times {\bm d}_m] \right ] \cr \cr
						= ({\bm E}_m \cdot  curl {\bm B}_m) {\bm B}_m +  ({\bm B}_m  \cdot {\bm d}_m) [{\bm E}_m \times {\bm B}_m] 
						- B^2_m [{\bm E}_m \times {\bm d}_m] = 0    		   ,  \cr  \cr
													  {D^2_m} (-div {\bm D}_m) =   {B_m^2} {{\bm D}_m} \cdot 
						  {\bm d}_m   - {\bm D}_m \cdot [{\bm B}_m  \times curl {\bm B}_m].
						 }						\end{eqnarray}

Stationary currents,  $- curl {\bm H}_m c/4\pi {\sqrt {g_{oo}}}
 \equiv - ({\bm B}_m \times {\bm d}_m + curl {\bm B}_m )  c/4\pi \neq 0$, in the vorticity monopole create local counteracting 
  ${\bm B}_m \times curl {\bm H}_m /4\pi {\sqrt {g_{oo}}}$ to prevent the Newtonian collapse  of continuous mass densities in their own gravitational forces ${\bm E}_m (- div  {\bm D}_m)  /4\pi {\sqrt {g_{oo}}}$. Static material states in  the continuous mass-energy integral correspond to the energy equipartition of \lq electric\rq{} and \lq magnetic\rq{} kinetic densities,   $D_m^2 = B_m^2$, initiated by the local vorticity.  Time-averaged vorticity forms metric space-matter with steady spatial displacement ${\bm D}_m$, averaged affine connections at $g_{oo} \neq const$, and averaged metric fields ${\bm E}_m$ for  local interactions with probe bodies. During measurements, the probe mass can only report changes in its (positive) energy integral, so that the observer can draw a conclusion about the accelerating (external) fields. In this way, the distributed vorticity of Cartesian material space creates its local displacement and, consequently, the continuous mass/charge densities, while the volumetric integral of these material densities can provide measurements (or practical observations). Despite vortex-magnetic monopoles stand behind the balancing creation of extended masses/charges in (5)-(6), only densest areas of such continuous charges, not monopoles with Poincar{\'e} radial pressure, can be probe bodies in direct measurements or mass-energy exchanges with experimental tools.           
	

Now let us analyze from (6) the static geodesic state of an isolated mass integral $m$ in its co-moving frame of reference ($ \gamma = 1 $) in order to find the equilibrium distribution in the mechanical charge ${{\sqrt G}} m = \int  {\sqrt {g_{oo}}}j^o d^3 x/c$ from 3-vector displacement ${\bm D}_m = \varphi_{\!_G} {\bm d}_m \equiv -  \varphi_{\!_G} {\bm \partial} ln{\sqrt {g_{oo}}}$  and energy equipartition $B^2_{m} = D^2_{m} = \varphi^2_{\!_G} d^2_m$ on the complementary degrees of freedom for the radial monopole and its charge.  
Taking ${\bm B}_m\cdot {\bm d}_m = {\bm E}_m \times  {\bm d}_m = 0$ form (5) for all static states, one can also put $curl {\bm B}_m = 0$ and  ${\bm d}_m \cdot [{\bm B}_m  \times curl {\bm B}_m ]= 0$ for the spherical symmetry of displacement 3-vectors  ${\bm D}_m(r) = \varphi_{\!_G} {d}_m(r) {\hat {\bf r}} $:
\begin {eqnarray}
     \cases {
		 - \varphi^2_{\!_G}{d}_m(r) div {d}_m(r){\hat {\bf r}} =    {B^2_m(r)}{d}_m(r) , \  - \partial_r [r^2 {d}_m(r)]  = r^2  d^2_m (r), \cr\cr
		  {d}_m(r) = -  r_o / r(r+r_o) = - \partial_r ln {\sqrt {g_{oo}(r)}}, \ g_{oo} = r^2/(r+r_o)^2 , \cr \cr
 \int {\sqrt {g_{oo}}} j^o d^3x/c ={\varphi_{\!_G} }\int_{o}^\infty d^2_m(r) d^3x/4\pi  =  {\varphi_{\!_G} }  r_o = {\sqrt G}m \ 
$(charge)$,
\cr\cr
B^2_m  = D^2_m = {\varphi^2_{\!_G} }r^2_o / r^2(r+r_o)^2,\  
 \int (D_m^2 + B^2_m)d^3x/8\pi =  m c^2 
 \ $(energy)$  
.  
	\cr	
						 }	
						\end{eqnarray}
								The static displacement  ${\varphi_{\!_G} }{d}_m(r){\hat {\bf r}} = 
								- {\hat {\bf r}}{\varphi_{\!_G} } [(1 / r) -1/(r+r_o)]$  
and the metric component $g_{oo} = r^2/(r+r_o)^2$, both found here from the 
Poincar{\'e} counter-force (6) of the vorticity monopole, were independently derived from the Machian relativistic analysis of GR fields \cite {Bul}.		

Unlike the Lorentz spherical electron with a finite radius, the mono-vertex distribution (7) with the radial monopole of inward vorticity (and its local pressure inferred by Poincare in 1906) extends over the entire Universe with the finite material density $m r_o/4\pi r^2 (r + r_o)^2$. Only half of the radial mass integral   $m$ belongs to a spherical volume with the dimensional scale $r_o = Gm/c^2$.  The other half of the isolated (radial) mass and associated inertial energy $mc^2/2$ are continuously distributed over micro, macro, and mega scales. Such a global superimposition of all \lq elementary\rq{} inhomogeneous densities in all spatial points of the system energy integral \cite {BulG} forms the nonlocal world continuum of correlated dynamical states with consistent densities, as in quantum mechanics of nonlocal material distributions.

	\section {Conclusions}					
		Changes of the mass-energy integral $mc^2 $ or mechanical charge $mc^2 / {\varphi_{\!_G} }$ of inertial systems can be measured in practice, say, during wave heating / cooling of examined bodies or after their inelastic collisions. Small positive mass-energy can be used as a probing entity to study external fields due to the locally observed velocity and acceleration of the densest regions of extended probe bodies.  Electric charges do not exist without inertial mass-energy as probing bodies should carry kinetic energy in the physical reality of local observations. But the physical limitations on the probing charge for performing local measurements do not deny the complementary coexistence of immeasurable monopoles of distributed vorticity, generating the inward Poincar{\'e} pressure regardless of measurements.

			
		It is significant that the continuous distribution of vorticity cannot be used in Cartesian physics as a probe charge. 
		But vorticity / magnetic fields always accompany	the extended  inertial/electric charge in a form of the  complementary monopole.  The vortex field of this monopole can be measured by probe charges according to the GR geodesics (3).  Since 1914, the latter associates any probe density with an independently existing mass-energy that is not part of the vortex system under study.

Today, the nonlocal organization of continuous mass-energy denies the model of independent or closed systems.
Instead, the joint material space of superimposed extended elements assumes their direct and unavoidable correlation on  micro-macro-mega levels in the world hierarchy of nonlocal distributions. Different geodesic equations (1) and (3), different mathematical descriptions and different dialectics should be applied to local observations and non-local self-organization in the world energy continuum, including inert and living subsystems.
Here Western dialectical thinking focused mainly on local measurements and observations in external fields according to (3) and the pragmatic predominance of Newtonian mechanics, while more ancient Eastern teachings traditionally focused on the non-local self-organization of correlated field energy in cosmos, described by its adaptive geodesics (1) for Cartesian space-matter.

The aforementioned analogy between mechanical and electrical geodesic self-forces in continuous densities (1) corresponds to the expected double unification of particles with fields and masses with electric charges. The latter can be consistently described by imaginary numbers, in order to restore the universal mechanical sign in Coulomb pair interactions and in the unphysical paradox of radiation self-acceleration  \cite {BulF}.

In addition to the locally observed dynamics of probing charges, there is the non-local organization of world kinetics energies with the local Poincar{\'e} pressure. 
This unobservable, inward regulation creates spatial displacements and observed densities of mass-energy. More precisely, elementary monopoles of continuously distributed vorticity stand behind the observed matter with inertial and electrical properties. 
Spatial displacements and measurable energies would never have occurred without the finite vorticity $ curl \gamma {\bm \beta} $ in the continuous space-matter of Plato, Aristotle and Descartes. Contrary to Newton gravitational energies, local tensions of  vorticity monopoles can lead to the novel description of metric material space in monistic terms of only positive (kinetic) energies.

Retaining the traditional approach to local measurements and retarded  exchanges between independent energy balances (the model of almost closed systems), the Cartesian worldview on matter-extension with coherently distributed vorticity can explain the instantaneous organization of inertial / charged densities in a nonlocal elementary system, including clouds of an isolated electron or stellar nebula. The material space physics sheds light on non-locally correlated quasars \cite {Ali}, the coordinated motion of  galaxies \cite {Mys}, the centuries-old stability of the Solar system contrary to the celebrated Laplace evaluations, and on many other pulsating and stable cosmic organizations of nonlocal distributions in Kant\rq{s} rational cosmology. The emerging experimental evidence of entangled macroscopic states may be a reason to include the nonlocal matter-extension with local Poincar{\'e} pressure in monistic kinetic mass-energy in the university curriculum alongside Newton\rq{s} dual alternative with kinetic and gravitational energies.

   
 


\end{document}